\documentclass[12pt]{article}
\usepackage{axodraw,mathrsfs}
\title{\bf Analysis of a model with a common source of
   CP violation}
\author{{\bf Biswajit Adhikary\footnote{biswajit.adhikari@saha.ac.in}}\\
   Saha Institute of Nuclear Physics,\\ 1/AF Bidhan  
        Nagar, Kolkata 700064, India }
\date{}

\begin{document}

\maketitle

\begin{abstract}
  We work in a model where all CP violating phenomena have a common
  source. CP is spontaneously broken at a large scale $V$ through the
  phase of a complex singlet scalar. An additional $SU(2)_L$ singlet
  vector-like down-type quark relates this high scale CP violation to
  low energy. We quantitatively analyze this model in the quark
  sector. We obtain the numerical values of the parameters of the
  Lagrangian in the quark sector for a specific ansatz of the $4\times
  4$ down-type quark mass matrix where the weak phase is generated
  minimally.  $Z \bar b b$ vertex will modify in presence of the extra
  vector-like down-type quark. From the experimental lower bound of
  the partial decay width $Z\rightarrow \bar b b$ we find out the
  lower bound of the additional down-type quark mass. Tree level
  flavor changing neutral current appears in this model due to the
  presence of the extra vector-like down-type quark.  We give the
  range of values of the mass splitting $\Delta m_{B_q}$ in
  $B^0_q-{\bar B}^0_q$ system using SM box, $Z$ mediating tree level
  and $Z$ mediating one loop diagrams together for both $q=d, s$.  We
  find out the analytical expression for $\Gamma_{12}^q$ in this model
  from standard box, $Z$ and Higgs mediated penguin diagrams for
  $B^0_q-{\bar B}^0_q$ system, $q=d,s$. From this we numerically
  evaluate the decay width difference
  $\left|\Delta\Gamma_{B_q}/\Gamma_{B_q}\right|$. We also find out
  the numerical values of the CP asymmetry parameters $a_J$ and
  $a_\pi$ for the decays $B^0_d\rightarrow J/\psi K_s$ and
  $B^0_d\rightarrow \pi^+ \pi^-$ respectively. We get the lower bound
  of the scale $V$ through the upper bound of the strong CP phase.

\end{abstract}

\section{Introduction}
\label{int}
CP violation is an important phenomena in the context of particle
physics and cosmology. CP violation is directly observed in the decays
of {\bf K} and {\bf B} mesons. The present experimental results
\cite{pdg} are consistent with the standard model(SM). The single
phase present in the Cabibbo, Kobayashi, Maskawa(CKM) mixing matrix is
responsible for this kind of CP violating phenomena. Present
observational data of neutrino oscillations prove that neutrinos are
massive. Minimal extension of the SM with neutrino mass generate
neutrino mixing similar to the quark mixing. So the phases in the
neutrino mixing matrix generate CP violating phenomena in the leptonic
sector. The strong CP violation comes from non-perturbative instanton
effects in the SM. This leads to the so called strong CP problem for
which various solutions have been proposed. The bound on the electric
dipole moment of neutron gives the bound on strong CP phase. In the
context of cosmology baryon asymmetry in the Universe(BAU) gives an
observational evidence for CP violation.  Decay of heavy Majorana
neutrino to lepton(both charged and neutral) and scalar(both charged
and neutral) generate lepton asymmetry which violates CP
\cite{fuk}.  One way of generation of baryon asymmetry is through
sphaleron mediated process from lepton asymmetry.

Different CP violating processes are not related in general. So the
question may arise whether it is possible to find a model where all
kinds of CP violation have a common origin. In fact there are a few
models \cite{Branco2,Achi}. We work in the model proposed by Branco,
Parada and Rebelo(BPR) \cite{Branco2}. CP breaks in this model
spontaneously through a $SU(3)_C\times SU(2)_L\times U(1)_Y$ singlet
complex scalar which gets a vacuum expectation value with nonzero
phase, $\frac{V}{\sqrt 2}\exp(i\alpha)$.  This large scale phase is
responsible for all kinds of CP violation. A vector-like down-type
quark and three generations of right handed neutrinos relate the large
scale phase to low energy CP violating phase in the quark sector and
the leptonic sector respectively. We quantitatively analyze this model
for the quark sector. In Section \ref{mod} we describe the model
\cite{Branco2} in detail. In Section \ref{anal} we calculate
parameters of the Lagrangian for a specific ansatz for the down-type
quark mass matrix, under the assumption that additional down-type
quark mass is larger than the standard down-type quark masses and
using the experimental range of values of the CKM parameters and the
standard down-type quark masses as the inputs. We take CKM matrix in
``standard'' parametrization \cite{param} for convenience. Presence of
a vector-like down-type quark generate flavor changing neutral
current(FCNC) at the tree level. It changes the flavor preserving
vertex as well.

 We see the effect in $Z\rightarrow \bar b b$ decay in
Section \ref{anal}. In this Section we get the lower bound of the
additional down-type quark mass from the experimental lower bound of
$Z\rightarrow \bar b b$ decay width. We also see the effect in $\Delta
m_{B_q}$ for $q=d,s$ of this additional down-type quark mass in the
same Section \ref{anal}. We also discuss the effect of this model on
$\Delta \Gamma_{B_q}$ in the same section for both $q=d,s$. We evaluate the expression
 of linearly  $U_{bd}$ dependent contribution to $\Gamma^q_{12}$ which
 is require to calculate the decay width differences. $U_{bd}$ is tree
 level coupling of $b$ and $d$ quarks with  $Z$ and also with neutral
  physical as well as fictitious scalar. We only consider the light
 Higgs contribution and disregard heavy scalar contribution. At last in this section we calculate the
numerical values of the CP asymmetry parameters $a_J$ and $a_\pi$ in
the decays $B^0_d\rightarrow J/\psi K_s$ and
  $B^0_d\rightarrow \pi^+ \pi^-$ respectively. Using the parametric solution and the extra
down-type quark mass bound to strong CP phase we get the lower bound
on the scale $V$ in Section \ref{cons}. Section \ref{conclu} contains
concluding remarks.

\section{The Model}
\label{mod}
The particle content of the standard model is
\begin{eqnarray}
\label{smp}
&&{q^\prime}_L=
\left(\begin{array}{c}u_L^\prime\\d^\prime_L\end{array}\right)
(3,2,\frac{1}{6}),\quad u_R^\prime(3,1,\frac{2}{3}),\quad 
 d^\prime_R(3,1,-\frac{1}{3}),\nonumber\\ 
 &&{\psi^l}^\prime_L=\left(\begin{array}{c}{\nu}^\prime_L\\l^\prime_L\end{array}\right)(1,2,-\frac{1}{2}), \quad l^\prime_R(1,1,-1),
\quad\Phi=\left(\begin{array}{c}\phi^+\\\frac{\phi_1+i\phi_2}{\sqrt{2}}\end{array}\right)(1,2,\frac{1}{2}).
\end{eqnarray}
$q_L^\prime$, ${\psi^l}^\prime_L$ and $\Phi$ are the standard
quark, lepton and scalar doublets respectively. Here $u_L^\prime$,
$u_R^\prime$, $d_L^\prime$ and $d_R^\prime$ are standard three generation left
handed up-type, right handed up-type, left handed down-type and right handed down-type quark
fields respectively. $l_L^\prime$, $l_R^\prime$ and $\nu_L^\prime$ are
standard three generation left handed lepton, right handed lepton and left
handed neutrino fields respectively.

The model of BPR
\cite{Branco2} has the same gauge symmetry as the SM, viz., $
SU(3)_C\times SU(2)_L\times U(1)_Y $. But in addition it has discrete symmetries $Z_4\times CP$. There are
also six
additional multiplets in this model along with the SM particles. One
is a totally gauge singlet complex scalar $S$. Others are the two
chiralities of an $SU(2)_L$ singlet
 vector-like down-type quark $D^\prime$, and three generation of
gauge singlet right handed neutrinos $N_R^\prime$.  These additional
multiplets have the following $ SU(3)_C\times SU(2)_L\times
U(1)_Y $ gauge group representations
\begin{eqnarray}
\label{adp}
D^\prime_R(3,1,-\frac{1}{3}),\quad D^\prime_L(3,1,-\frac{1}{3}),\quad N^\prime_R(1,1,0),\quad S(1,1,0).
%\widetilde\Phi=\left(\begin{array}{c}\frac{\phi_1-i\phi_2}{\sqrt{2}}\\-\phi^-\end{array}\right)
%(1,2,-\frac{1}{2}),\nonumber\\
%&&\quad S(1,1,0), \quad S^*(1,1,0).
\end{eqnarray}
The primes on the fermion fields in Eq.\ (\ref{smp}) and Eq.\
(\ref{adp}) imply that they do not necessarily have
definite mass.  Under $Z_4$ the fields transform as,
\begin{eqnarray}
\label{ftrans}
 S\rightarrow {-S},\quad D^\prime\rightarrow{-D}^\prime, 
\quad{\psi^l}^\prime_L\rightarrow i{\psi^l}^\prime_L,
\quad l^\prime_R\rightarrow il^\prime_R,\quad N^\prime_R\rightarrow
iN^\prime_R\ 
\end{eqnarray}
while all the other fields remain invariant.
  The Yukawa interactions in this model are,
\begin{eqnarray}
\label{ylagq}
-\mathscr L^{Y}_{q}&=&{{\bar{q^\prime}}_L} Y^d\Phi d^\prime_R+
{\bar{q^\prime}_L} Y^{u}\widetilde\Phi
{u_R^\prime}
+(f_qS+{f_q}^\prime S^*){\bar D}^\prime_L{d^\prime_R}
+\mu_0{\bar D}^\prime_L{D^\prime_R}+h.c. 
\end{eqnarray}
in the quark sector and
\begin{eqnarray}
\label{ylagl}
-\mathscr L^{Y}_{l}&=&{{\bar{\psi}}}^{l^\prime}_LY^l\Phi l^\prime_R+
{\bar{\psi}}^{l^\prime}_L Y^{\nu}\widetilde\Phi
{N^\prime_R}
+{{N^\prime}}^T_RC^{-1}(f_{\nu}S+{f_\nu}^\prime S^*) {N^\prime_R}
+h.c.
\end{eqnarray}
 in the
leptonic sector, where $\widetilde\Phi = i\sigma_2\Phi^*$ and $S^*$ is
the conjugate field of $S$. $Y^u$, $Y^d$ in Eq.\
(\ref{ylagq}) and $Y^l$ in Eq.\
(\ref{ylagl}) are the standard $3\times 3$ Yukawa coupling
matrices. $f_q$, $f_q^\prime$ in Eq.\ (\ref{ylagq}) are the nonstandard $1\times 3$ Yukawa
coupling matrices. $Y^\nu$, $f_\nu$ and $f^\prime_\nu$ in Eq.\
(\ref{ylagl}) are also nonstandard $3\times 3$ Yukawa coupling
matrices. Without loss of generality we choose the basis of the fields such that
$Y^u$ is diagonal with the real positive entities. $C$ in Eq.\ (\ref{ylagl}) is
 a $4\times 4$ matrix which relate any charge conjugated four spinor field to
 its original field, $\psi^c=\gamma_0 C\psi^*$. 
 The Lagrangian ${\mathscr L}^Y_q$ in Eq.\ (\ref{ylagq}) has bare mass
term of $D^\prime$ where $\mu_0$ is the bare mass parameter. Presence of
this term does not violate any symmetry of the Lagrangian.  All the
couplings of the Lagrangian of this model are real due to CP
symmetry. The most general gauge, CP and $Z_4$ invariant scalar
potential \cite{Branco1} of this model is 
\begin{eqnarray}
\mathscr{V}&=&-\mu^2\Phi^\dagger\Phi+\lambda(\Phi^\dagger\Phi)^2+SS^*(a_1+b_1SS^*)+(S^2+{S^*}^2)(a_2+b_2SS^*)\nonumber\\&&+b_3(S^4+{S^*}^4)+(\Phi^\dagger\Phi)\{c_1(S^2+{S^*}^2)+c_2SS^*\}.
\label{scalp}
\end{eqnarray}
Because of the terms $a_2$, $b_2$, $b_3$ and $c_1$ the field $S$ can
acquire a complex vacuum expectation value (VEV) :
$\left<S\right>=\frac{V}{\sqrt{2}}\exp(i\alpha)$. This VEV breaks CP
and $Z_4$ spontaneously\footnote{Spontaneous breaking of any discrete
  symmetry generates cosmological domain wall problem. Solution of
  this problem goes beyond the scope of this article.}. The
$SU(2)_L\times U(1)_Y$ symmetry spontaneously breaks when the scalar
$\Phi$ takes VEV,
$\left<\Phi\right>=\left(\begin{array}{c}0\\\frac{v}{\sqrt{2}}\end{array}\right)$.
It is natural to think that $V$ $\gg$ $v$ due to the fact that the
scalar $S$ is a gauge singlet.
\section{Quantitative analysis in the hadronic sector}
\label{anal}
The VEV of $\Phi$ and $S$ break the $SU(3)_C\times SU(2)_L\times U(1)_Y\times Z_4\times CP$
 symmetry down to  $SU(3)_C\times U(1)_Q\times Z_2$. This generates mass terms. Mass part of
the Lagrangian in the quark sector is,
\begin{eqnarray}
\label{lagm}
-{\mathscr L}^m_q={\bar d}^\prime_L m_d^0 d^\prime_R+
{\bar{u}}_L^\prime m^0_u
u_R^\prime
+M_D^0{\bar D}^\prime_L d^\prime_R
+\mu_0{\bar D}^\prime_L D^\prime_R+h.c.
\end{eqnarray}
where,
\begin{eqnarray}
\label{mmex}
 m^0_u=\frac{v}{\sqrt 2}Y^u,\quad m_d^0=\frac{v}{\sqrt
2}Y^d
\end{eqnarray}
and
\begin{eqnarray}
\label{Dmm}
 M_D^0=\frac{V}{\sqrt 2}\{f_q\exp(i\alpha)+{f_q}^\prime
\exp(-i\alpha)\}.
\end{eqnarray}
Since we are in a basis where $Y^u$ is
diagonal with the real positive entities, the up-type quark mass matrix $m^0_u$ will also be
diagonal, $m^0_u=$ diag($m_u$, $m_c$, $m_t$). So the up-type fields
$u^\prime$ 
are physical. We will call them $u$(unprimed) for future
reference. This is our convention. The down-type quark
mass term from Eq. ($\ref{lagm}$) can be written by the following form
\begin{eqnarray}
\label{mform}
-\mathscr{L}^m_d=\left(\begin{array}{cc}{\bar{d}^\prime_L} &
    {\bar{D^\prime}_L}\end{array}\right)\left(\begin{array}{cc}\{{m_d^0}\}_{\bf{3\times 3}} & \{0\}_{\bf {3\times 1}}\\\{{M_D^0}\}_{\bf {1\times 3}} & \{{\mu_0}\}_{\bf {1\times 1}}\end{array}\right)\left(\begin{array}{c}{d}^\prime_R \\ {D^\prime}_R\end{array}\right).
\end{eqnarray}
The above mentioned  $4 \times 4$ down-type quark mass matrix is denoted by $\mathscr
{M}$.  $\mathscr {M}$ can be diagonalized by bi-unitary transformation,
\begin{eqnarray}
\label{diagm}
\mathscr{U}^\dagger\mathscr{M}\mathscr{U}^\prime=\left(\begin{array}{clcr}{\bar m}_d
 & 0\\0 &
M_D\end{array}\right),
\end{eqnarray}
where ${\bar m}_d=$ diag($m_d$, $m_s$, $m_b$) and $M_D$ is the mass of
additional down-type quark. Let us consider the forms of $\mathscr{U}$ and $\mathscr{U}^\prime$ as follows
\begin{eqnarray}
\label{uform}
\begin{array}{cc}\mathscr{U}=\left(\begin{array}{cc}K_{\bf{3\times
          3}} & R_{\bf {3\times 1}}\\ S_{\bf {1\times 3}} & T_{\bf
        {1\times 1}}\end{array}\right), &
 \mathscr{U}^\prime=\left(\begin{array}{cc}K^\prime_{\bf{3\times 3}} & R^\prime_{\bf {3\times 1}}\\ S^\prime_{\bf {1\times 3}} & T^\prime_{\bf {1\times 1}}\end{array}\right)\end{array}.
\end{eqnarray}
Relations of the physical basis(unprimed) of the down-type quark fields  to the
original basis are
\begin{eqnarray}
\label{btm1}
 \left(\begin{array}{c}{ d}_L\\{ D}_L\end{array}\right)=\left(\begin{array}{clcr}K^\dagger & S^\dagger\\R^\dagger &
T^*\end{array}\right)\left(\begin{array}{c}{ d^\prime}_L\\{
  D^\prime}_L\end{array}\right)
\end{eqnarray}
and
\begin{eqnarray}
\label{btm2}
\left(\begin{array}{c}{ d}_R\\{ D}_R\end{array}\right) =\left(\begin{array}{clcr}{K^\prime}^\dagger & {S^\prime}^\dagger\\{R^\prime}^\dagger &
{T^\prime}^*\end{array}\right)\left(\begin{array}{c}{ d^\prime}_R\\{ D^\prime}_R\end{array}\right).
\end{eqnarray}
Clearly $\mathscr{U}$ diagonalizes $\mathscr {M}\mathscr {M}^\dagger$:
\begin{eqnarray}
\label{mds}
\mathscr{U}^\dagger\mathscr{M}\mathscr{M}^\dagger\mathscr{U}=\left(\begin{array}{clcr}{\bar m}_d^2
 & 0\\0 &
M_D^2\end{array}\right).
\end{eqnarray}
This gives the approximate relation \cite{Branco2,Branco1}
\begin{eqnarray}
\label{emd}
\mathscr{H}={K}{\bar m}_d^2K^{-1}
\end{eqnarray}
where
\begin{eqnarray}
\label{efm}
\mathscr{H}={m_d^0}{{m_d^0}}^\dagger-\frac{{m_d^0}{M_D^0}^\dagger
M_D^0{{m_d^0}^\dagger}}{M_D^2}
\end{eqnarray}
 is the effective
 mass matrix square for the standard down-type quarks under the  assumption  
\begin{eqnarray}
\label{assm}
\frac{{{\bar{m}_d}}^2}{M_D^2}\ll 1,
\end{eqnarray}
and with
\begin{eqnarray}
\label{Dqm}
M_D^2\approx M_D^0{M_D^0}^\dagger+\mu_0^2.
\end{eqnarray}
$K$ is the CKM matrix and it is approximately unitary for the
hermiticity of $\mathscr{H}$. The other blocks of $\mathscr{U}$ in
Eq.\ (\ref{uform}), under the assumption in Eq.\ (\ref{assm}) obtain
the forms \cite{Branco1}
\begin{eqnarray}
\label{ulbs}
R\approx\frac{{{m_d^0}}{M_D^0}^\dagger}{M_D^2},\qquad T\approx
 1
\end{eqnarray}
and 
\begin{eqnarray}
\label{sfm}
S\approx -\frac{M_D^0{{{m_d^0}}^\dagger}K}{M_D^2}.
\end{eqnarray}
The constraints of unitarity of the matrix $\mathscr{U}$ defined in Eq.\
(\ref{uform}), imply the relations
\begin{eqnarray}
\sum_j\left|K_{ij}\right|^2=1-\left|R_{i}\right|^2
\label{weakuniv}
\end{eqnarray}
for all generations $i$, $i=u, c, t$. This implies that the sum of all
couplings of any of the up-type quarks to all the standard down type
quarks is not same for all generations. Presence of
$\left|R_{i}\right|^2$ term in the right hand side of Eq.\ 
(\ref{weakuniv}) breaks the weak universality. Smallness of the 2nd
term in the right hand side of Eq.\ (\ref{weakuniv}) due to the
assumption in Eq.\ (\ref{assm}) suggests that unitarity of CKM matrix
$K$ and weak universality hold only approximately in this model.

Here the CKM phase can be generated through the second term of
$\mathscr{H}$ in Eq.\ (\ref{efm}). Note that this term is not
suppressed by the scale $V$ because $M_D^0$ and $M_D$, as we have seen
from Eq.\ (\ref{Dmm}) and Eq.\ (\ref{Dqm}), both have dependence on
the scale $V$.

Now we try to analyze this model quantitatively. Since up-type quark mass matrix is diagonal, the remaining parts of ${\mathscr L}^m_q$ in Eq.\
(\ref{lagm}) have $18$ real
parameters. These parameters are $9$ of ${m_d^0}$, $3$ of  $f_q$, $3$
of  $f_q^\prime$ and other $3$ are
 $\alpha$, $V$ and $\mu_0$. The matrix $\mathscr{M}$ in Eq.\
 (\ref{diagm}) and hence $\mathscr{H}$ in Eq.\
 (\ref{efm}) contains all these
parameters. The question may arise whether it is possible to find
realistic solution of the Eq.\ (\ref{emd}) for the entire range of 
 values
of  the CKM
parameters and the masses of the standard down-type quarks. Since $\mathscr{H}$ is
hermitian,
there are $9$ independent coupled equations consisting of these $18$ 
parameters as variables. The reduction of the number of variables can be
obtained as the following way.  First we assume that ${m_d^0}$ is
symmetric. Then we note that $M_D$ and $M_D^0$ occur in Eq.\
(\ref{efm}) in the combination
\begin{eqnarray}
\label{fdm}
F_D=\frac{M_D^0}{M_D},
\end{eqnarray}
which can be parametrized as follows 
\begin{eqnarray}
F_D^T=\left(\begin{array}{c}{f_1\exp(i\alpha)+f_1^\prime \exp(-i\alpha)}\\{f_2\exp(i\alpha)+f_2^\prime 
\exp(-i\alpha)}\\{f_3\exp(i\alpha)+f_3^\prime \exp(-i\alpha)}\end{array}\right),
\end{eqnarray}
where $f_i=\frac{V}{\sqrt{2}M_D}(f_q)_i$ and
$f^\prime_i=\frac{V}{\sqrt{2}M_D}(f_q^\prime)_i$. Now if we consider
$\bf{f}=\bf{f^\prime}$, the matrix $\mathscr{H}$ in Eq.\ (\ref{efm})
becomes real symmetric. This implies that the CKM phase is
zero. Now let us see minimally how we can generate a non zero CKM phase. To this end, we
take the ansatz
\begin{eqnarray}
\label{fdf}
f_1=f_1^\prime=f\nonumber\\
f_2=f_2^\prime=\beta f\nonumber\\
f_3=f+\Delta f\nonumber\\
f_3^\prime=f.
\end{eqnarray}
\begin{table}
\begin{center}
\begin{tabular}{|c|c|c|c|}
\hline
Type of Input Parameters & Specification & Experimental Range & Random Choice \\                   
\hline
CKM Parameters & $\sin\theta_{12}$ & $0.2227$ to $0.2259$ & $0.2238$\\ \cline{2-4}
 & $\sin\theta_{13}$ & $0.0032$  to  $0.0042$ &   $0.0037$\\ \cline{2-4}
 & $\sin\theta_{23}$ & $0.0398$  to  $0.0428$ & $0.0398$  \\ \cline{2-4}
 & $\delta_{CKM}$ & $46^\circ$  to  $74^\circ$ & $73.6818^\circ$  \\
\hline
Mass Parameters & $m_d$ & $4.0$   to   $8.0$ $MeV$ & $5.9171$ $MeV$ \\ \cline{2-4}
 & $m_s$ & $80.0$   to   $130.0$ $MeV$ & $81.6310$ $ MeV$  \\ \cline{2-4}
 & $m_b$ & $4.1$   to   $4.4$ $GeV$ &  $4.3358$ $GeV$  \\ 
\hline
\end{tabular}\\
\caption[] {\label{ckmas}Here $\theta_{ij}$'s are the CKM angles and $\delta_{CKM}$ is
CKM phase in ``standard'' parametrization of CKM matrix \cite{param}. Experimental range of the $\sin$ of the CKM angles, CKM
phase and the standard down-type quark
masses are taken from PDG \cite{pdg}. In \cite{pdg} the values quoted
for mass of $m_d$, $m_s$ are measured at the scale $2$ GeV and
$m_b\equiv \bar{m}_b(\bar{m}_b)$ in the $\overline{\rm MS}$
scheme. We randomly choose a set of the above
inputs from their experimental range.}  
\end{center}
\end{table}
So ultimately we have 10 variables which are $f$, $\Delta f$,
$\alpha$, $\beta$ and six elements of ${m_d^0}$. So we are in a
position where we can find the parameters of the Lagrangian from
Eq.\ (\ref{emd}).
 We choose the values of the CKM parameters
and the masses of the standard down-type quarks from their experimental
range as in Table \ref{ckmas} . Using these values in the right hand side  of the Eq. (\ref{emd}) 
we solve this equation for different values of
$\beta$.  
We observe that there is a range of the values of $\beta$, $-3.2<\beta<2.12$, where Eq.\ (\ref{emd}) fails to give real solutions
for the parameters of $\mathscr{H}$  for the entire range of values of the CKM
parameters and the standard down-type quarks masses as in Table \ref{ckmas}. We get
the real
solutions outside this range of  values of $\beta$. 

To get the essence of the solutions let us show that how we 
have 
proceeded to reach the goal. First we write ${m_d^0}$ and $F_D$ in the
following forms: 
\begin{eqnarray}
\label{matf}
\begin{array}{cc}{{m_d^0}=v\left(\begin{array}{ccc}
x_1 & x_2 & x_3 \\
x_2 & x_4 & x_5 \\
x_3 & x_5 & x_6 
\end{array}\right)} & \qquad
{{{F_D}^T}=\sqrt{x_7}\left(\begin{array}{ccc}
1 \\
\beta \\
x_8+ix_9  
\end{array}\right)}\end{array}.
\end{eqnarray}
 Here $x_1$ to $ x_6$ are six independent dimensionless mass
parameters of ${m_d^0}$. $x_7$ to $x_9$ are three
independent dimensionless parameters of $F_D$. We have written $F_D$
 in Eq.\ (\ref{matf}) with a pre-factor $\sqrt{x_7}$. This makes  $x_7$
 dependence of 
 Eq.\ (\ref{emd}) linear. The $x_7$ to $x_9$ are related to
$f$, $\Delta f$ and $\alpha$ in the following way
\begin{eqnarray}
\label{fxr}
x_7=4f^2\cos^2{\alpha},\quad x_8=1+\frac{\Delta f}{2f},\quad
x_9=\frac{\Delta f\tan{\alpha}}{2f}.
\end{eqnarray}
For every set of inputs we have four sets of solutions. These can be divided into two pairs of sets, where within a pair, the two solutions are related only by change of signs for $x_1$ to $x_6$. The solutions of different
pairs have the values of the same order of magnitudes. We provide two
 distinct sets of solutions in Table \ref{xsol} for a set of
inputs in the
rightmost column of Table \ref{ckmas} with $\beta=3.5$ and $\beta=-3.5$. Inputs  are chosen randomly from their
experimental range in Table \ref{ckmas}.
\begin{table}
\begin{center}
\begin{tabular}{|c|c|c|c|c|}
\hline
Variables & \multicolumn{2}{|c|}{Results For $\beta$=$3.5$} & \multicolumn{2}{|c|}{Results For $\beta$=$-3.5$}\\\cline{2-5}
 & Set No. 1  & Set No. 2 &  Set No. 1  & Set No. 2 \\     
\hline
$x_1$ & $ 0.000162841$ & $0.000168356$  & $-0.000008467$ & $-0.0000154413$\\
\hline
$x_2$ & $0.000301087$ & $0.000320389$  & $0.000298491$ & $0.000322901$\\ 
\hline
$x_3$ & $-0.000255079$ & $0.000229067$  & $-0.000320499$ & $0.000291728$\\ 
\hline
$x_4$ & $-0.000122682$ & $-0.0000551233$ & $0.000131765$ & $0.0000463298$ \\ 
\hline
$x_5$ & $-0.000845869$ & $0.000848642$  & $0.00107484$ & $-0.00106796$\\
\hline
$x_6$ & $-0.0212162$ & $0.0212857$  & $0.026958$ & $-0.0267854$\\ 
\hline
$x_7$ & $0.0488582$ & $0.0539958$  & $0.0289893$ & $0.0353266$\\ 
\hline
$x_8$ & $-2.53447$ & $2.14896$  & $4.41581$ & $-3.72177$\\ 
\hline
$x_9$ & $-0.849496$ & $0.766168$  & $-1.28386$ & $1.06035$\\
\hline
\end{tabular}
\caption{\label{xsol}Two distinct sets of solutions of the Eq.\ (\ref{emd}) for
the inputs of the rightmost column of Table \ref{ckmas} with $\beta=\pm 3.5$.} 
\end{center}
\end{table}

Now let us see that what are the outcomes of these
solutions. ${m_d^0}$ becomes explicitly known. $F_D$ also
becomes  known meaning that Yukawa parameters related to the
coupling of the standard down-type quarks, extra vector like down-type quark and the singlet 
scalar also become known upto a factor $M_D$/$V$. Using the relations
 in Eq.\ (\ref{fxr}) we obtain the values of $f$,
$\Delta f$ and $\alpha$  in Table \ref{lpar} for the different solutions in Table \ref{xsol}.
\begin{table}
\begin{center}
\begin{tabular}{|c|c|c|c|c|}
\hline
Quantities & \multicolumn{2}{|c|}{Results For $\beta$=$3.5$} & \multicolumn{2}{|c|}{Results For $\beta$=$-3.5$}\\\cline{2-5}
 & Set No. 1  & Set No. 2 &  Set No. 1  & Set No. 2 \\     
\hline
$f$ & $\pm 0.113667$ & $\pm 0.139648$ & $\pm 0.0909458$  & $\pm 0.0963174$ \\
\hline
$\Delta f$ & $\mp 0.803504$ & $\pm 0.3209$  & $\pm 0.621307$& $\mp 0.909577$ \\ 
\hline
$\alpha$ & $13.5145$ & $33.6967$  & $-20.5991 $& $-12.6567$ \\ 
\hline
$F_DF_D^\dagger$ & $0.996472$ & $0.996495$  & $0.997163 $& $0.997127$ \\ 
\hline
\end{tabular}
\caption{\label{lpar}Values of $f$, $\Delta f$,
  $\alpha$ and $F_DF_D^\dagger$ for the solution in Table \ref{xsol} using Eq.\ (\ref{fxr}).} 
\end{center}
\end{table}
The $R$ block of ${\mathscr{U}}$ as in Eq. ($\ref{uform}$) is the
coupling of charged current interactions among up-type quarks $u$ and $D$.
$S^\dagger S$ is a matrix which generates FCNC for the standard down-type quarks. Using Eq.\
(\ref{fdm}), we can rewrite Eq.\ (\ref{ulbs}) and Eq.\ 
(\ref{sfm}) as 
\begin{eqnarray}
R&=&\frac{{{m_d^0}}{F_D}^\dagger}{M_D}\nonumber\\
S&=& -\frac{F_D{{{m_d^0}}^\dagger}K}{M_D}.
\label{rsf}
\end{eqnarray}
The numerator of $R$ and $S$ in above Eq.\ (\ref{rsf}) become known
from the numerical solution. Their
forms for the Set-1 solution with $\beta=3.5$ in Table \ref{xsol} are
\begin{eqnarray}
\label{ofdp}
\begin{array}{cc}{R=\frac{1}{M_D}\left(\begin{array}{c}
0.101309-i0.0117825\\
0.109596-i0.0390722\\
2.74902-i0.980013    
\end{array}\right)} & {S^T=\frac{1}{M_D}\left(\begin{array}{c}
-0.0993492-i0.000938182\\
-0.0228038-i0.000215232 \\
-2.75133-i0.980436  
\end{array}\right)}\end{array}.
\end{eqnarray}
Now let us observe two parts of the $\mathscr{H}$ matrix in Eq.\ 
(\ref{efm}). Using the Set-1 solution for $\beta=3.5$ in Table
\ref{xsol} we have
\begin{eqnarray}
\label{fpat}
{m_d^0}{{m_d^0}}^\dagger=\left(\begin{array}{ccc}
0.0110282 & 0.0137888 & 0.309574\\
0.0137888 & 0.0496956 & 1.08766\\
0.309574 & 1.08766 &  27.287
\end{array}\right)
\end{eqnarray}
and
\begin{eqnarray}
\label{spat}
&&\frac{{m_d^0}{M_D^0}^\dagger
M_D^0{{m_d^0}}^\dagger}{M_D^2}=\nonumber\\&&\left(\begin{array}{ccc}
0.0104023 & 0.0115634+i0.00266705 & 0.290048+i0.0668936 \\
0.0115634-i0.00266705 & 0.0135379 & 0.339572-i5.14869\times 10^{-6}\\
0.290048-i0.0668936 & 0.339572+i5.14869\times 10^{-6}  & 8.51754
\end{array}\right).\nonumber\\
\end{eqnarray}
It is interesting to see that two parts of $\mathscr{H}$ contributes roughly
equally for many elements. The second part of $\mathscr{H}$ in Eq.\ (\ref{spat}) is new physics
term and it is not suppressed. This 
feature is independent of the inputs we are giving. Another
input independent feature is that the contribution of bare part to
the mass of $D$ quark $M_D$ in Eq.\ (\ref{Dqm}) is smaller than the Yukawa
part. The Yukawa part $M_D^0{M_D^0}^\dagger/{M_D^2}=F_DF_D^\dagger$ contributes
 approximately $99.64\%$ to $M_D^2$ as in Table \ref{lpar} for the solutions in
 Table 2 with $\beta=3.5$. The contribution for the solutions
 $\beta=-3.5$ is nearly $99.71\%$ as in Table \ref{lpar}. Although bare
mass term is gauge invariant, its contribution to the mass of
$D$ quark is
subdominant. One conclusion we may draw from here that $M_D$
should be less than $V$ for the validity of perturbation theory. Now let us see that how we can get more information about
$M_D$ from $Z\rightarrow b\bar b$ decay and how $M_D$ value affects $B^0_d-{\bar{B}}^0_d$ mixing. 

The presence of extra vector quark generates FCNC at the tree level. It also changes the flavor
preserving neutral current. The interaction of $Z$ and the standard
down-type quark is
\begin{eqnarray}
\label{zdd}
\mathscr{L}_{NC}^d=-\frac{g}{2\cos{\theta_W}}\sum_{i,j=1}^3{\bar
d_i}\gamma_\mu(g_V-\gamma_5 g_A)_{ij}d_j Z^\mu
\end{eqnarray}
where $d_1=d$, $d_2=s$, $d_3=b$, 
\begin{eqnarray}
(g_V)_{ij} = g_{V}^{SM}\delta_{ij}-I_3^dU_{ij}
\label{vcoup}
\end{eqnarray}
and
\begin{eqnarray}
(g_A)_{ij} = g_{A}^{SM}\delta_{ij}-I_3^dU_{ij}
\label{acoup}
\end{eqnarray}
where  
\begin{eqnarray}
U=S^\dagger S
\label{udef}
\end{eqnarray}
and the tree level couplings in SM are
\begin{eqnarray}
g_{V}^{SM}&=&I_3^d-2Q\sin^2{\theta_W},\\\nonumber
g_{A}^{SM}&=&I_3^d
\label{stncop}
\end{eqnarray}
with $I_3^d=-1/2$, $Q=-1/3$.  The flavor preserving vertex appreciably
changes by loop corrections. We want to see the effect of loop
correction along with the new contribution of $U_{bb}$ dependent terms
in $Z\rightarrow b\bar b$ process \cite{gb}. The partial decay width
of $Z\rightarrow b\bar b$ in terms of effective coupling constants
$g_{V_{\rm eff}}^b$ and $g_{A_{\rm eff}}^b$ is \cite{hol}
\begin{eqnarray}
\label{dwt}
\Gamma_{Z\rightarrow b\bar b}=\frac{\sqrt{2}G_FM_Z^3}{4\pi}&&[(1-4y)^{\frac{1}{2}}\{
    |g_{V_{\rm eff}}^b |^2(1+2y)+ |g_{A_{\rm eff}}^b
  |^2(1-4y)\}
\times (1+\delta_{QED})\nonumber\\&&+(
    |g_{V_{\rm eff}}^b |^2+ |g_{A_{\rm eff}}^b
  |^2)\delta_{QCD}]+\Delta_{QCD}(y)
\end{eqnarray}
where $y=m_b^2/M_Z^2$, QED correction $\delta_{QED}$ and factorizable
QCD correction $\delta_{QCD}$ are
\begin{eqnarray}
\delta_{QED}&=&3\alpha Q^2/{4\pi}, \nonumber\\
\delta_{QCD}&=&(\alpha_s/\pi)+1.41(\alpha_s/\pi)^2-12.8(\alpha_s/\pi)^3-\alpha\alpha_s Q^2/{4\pi^2},
\label{qcqe}
\end{eqnarray}
and the effective couplings are
\begin{eqnarray}
g_{V_{\rm eff}}^b&=&\sqrt{\rho_b}
(I_3^b-I_3^bU_{bb}-2Q\sin^2\theta_{W}\kappa_b)\\\nonumber
g_{A_{\rm eff}}^b&=&
\sqrt{\rho_b}(I_3^b-I_3^bU_{bb}).
\label{efcoup}
\end{eqnarray}
Using general expression for $\rho_b$ and $\kappa_b$
\cite{hol,fles,adv} with three loop QCD correction and two loop
electroweak correction, we obtain $0.9935\le\rho_b\le 0.9941$ and
$1.0341\le\kappa_b\le 1.0382$ for SM using experimental range of
values of $\alpha_s (M_Z)$, $m_t$, $M_W$, $M_Z$ and $m_b$ \cite{pdg}.
Hence, we get that the SM prediction is $0.37655 \le
\Gamma_{Z\rightarrow b\bar b} \le 0.37869$ GeV. $\Delta_{QCD}(y)$ in
Eq.\ (\ref{dwt}) is $b$ quark mass dependent QCD correction. We
disregard here small non-factorizable QCD correction. New physics in
the form of $U_{bb}$ can decrease $g_{V_{\rm eff}}^b$, $g_{A_{\rm
    eff}}^b$ and hence the decay width $\Gamma_{Z\rightarrow b\bar
  b}$, since by definition $0\le U_{bb}\le 1$. Experimental data at the
$1\sigma$ level given $0.37593 \le \Gamma_{Z\rightarrow b\bar b} \le
0.37912$ GeV \cite{pdg}. If we take the SM contribution at the
theoretical lower limit $0.37655$ GeV, we would then need
\begin{eqnarray}
U_{bb}\le 7.204\times
10^{-4}
\label{ubbb}
\end{eqnarray}
in order that the new physics contributions do not violate the
experimental bound.
From the definition of matrix $U$ in Eq.\ (\ref{udef}) and using the
form of $S$ in Eq.\ (\ref{rsf}) we see that $U$ is proportional to
$1/M_D^2$ which is unknown.  Hence we can use the upper bound on
$U_{bb}$ to obtain a lower bound on $M_D$. We provide these bounds in
Table \ref{mdtv} for different values of $\beta$.  The upper bound of
the value of the other elements of $U$ become known from the lower
bound on $M_D$.  For the different sets of solution $M_D$ values are
approximately same for a given set of inputs. The least value of $D$
quark mass $M_D\approx 40$ GeV for the given range of $\beta$ in the
Table \ref{mdtv}. The maximum value among the elements in matrix
$\bar{m}_d$ is $b$ quark mass which is nearly $4$ GeV. So the
assumption $\bar{m}_d^2/M_D^2$$\ll 1$ in Eq.\ (\ref{assm}) is not too
bad at all.  This assumption is better for smaller $\beta$ region
where $M_D$ has to be larger.

We now want to see the effect in flavor changing process.  $U$
dependent extra piece in $g_V$ and $g_A$ in Eq.\ (\ref{vcoup}) and
Eq.\ (\ref{acoup}) are responsible for the FCNC.  These generate tree
level $Z$ mediating $B^0_q-{\bar B}^0_q$ mixing, where $q=d, s$.  The
experimental value of the mass splitting $\Delta m_{B_d}$ in
$B^0_d-{\bar{B}}^0_d$ system is well known, $\Delta m_{B_d}^{\rm
  ex}=(3.304\pm .046)\times 10^{-10}$ MeV \cite{pdg}. For
$B^0_s-{\bar{B}}^0_s$ system only lower bound exists, $\Delta
m_{B_s}>94.8\times 10^{-10}$ MeV at $95\%$ CL \cite{pdg}.  The
contribution of SM box \cite{bura1}, $Z$ mediated tree diagrams
\cite{nir,branco3} and $Z$ mediated one loop diagrams together in
$\Delta m_{B_q}$ has been explictly calculated by Barenboim and
Botella \cite{baren}\footnote{Although the authors in \cite{baren}
  calculated $\Delta m_{B}$ for $B^0_d-{\bar{B}}^0_d$ system, it can
  be easily extended to $B^0_s-{\bar{B}}^0_s$ system also.}, who give
off diagonal term of $B^0_q-{\bar{B}}^0_q$ mixing matrix $M_{12}^q$ as
\begin{eqnarray}
M_{12}^q=M_{12}^{qSM}\Delta_{bq}^{*}
\label{bdme}
\end{eqnarray}
where
\begin{eqnarray}
\label{bmix}
M_{12}^{qSM}=\frac{G_F^2M_W^2\eta_{B}
  m_{B_q}(B_{B_q}f_{B_q}^2)}{12\pi^2}\bar E(x_t)(\xi^{q*}_t)^2,
\end{eqnarray}
and 
\begin{eqnarray}
\Delta_{bq}=1-a(U_{bq}/\xi^q_t)-b(U_{bq}/\xi^q_t)^2.
\label{delbd}
\end{eqnarray}
Here $a$ and $b$ are
\begin{eqnarray}
a&=&4\frac{\bar C(x_t)}{\bar E(x_t)}\nonumber\\
b&=&\frac{4\pi\sin ^2\theta _W}{\alpha}\frac{1}{\bar E(x_t)}
\label{dab}
\end{eqnarray}
where
\begin{eqnarray}
\bar E(x_t)&=&\frac{-4x_t+11x_t^2-x_t^3}{4(1-x_t)^2}+\frac{3x_t^3\ln
  x_t}{2(1-x_t)^3}\nonumber\\
\bar C(x_t)&=&\frac{x_t}{4}\left[\frac{4-x_t}{1-x_t}+\frac{3x_t\ln
  x_t}{(1-x_t)^2}\right].
\label{ecd}
\end{eqnarray}
The parameters involved in $ M_{12}^{qSM}$ in Eq.\ (\ref{bmix})
$m_{B_q}$ and $f_{B_q}$ are mass and decay constant of $B^0_q$ meson
respectively. $B_{B_q}$ is Renormalization Group invariant parameter.
$\eta_{B}$ is QCD factor.  Its value is nearly $0.55$ \cite{bura}.
$m_{B_d}=5279.4\pm0.5$ MeV and $m_{B_s}=5369.6\pm 2.4$ MeV \cite{pdg}.
$x_t=m_t^2/M_W^2$. $\xi^q_t=V^*_{tb}V_{tq}$, $q=d,s$. The uncertainty in the calculation come from $\sqrt
{B_{B_q}}f_{B_q}$. $\sqrt {B_{B_d}}f_{B_d}=221\pm 28^{+0}_{-22}$ MeV
and $\sqrt {B_{B_s}}f_{B_s}=255\pm 31$ MeV which are obtained from the
lattice QCD calculations in \cite{hurt}.  Mass splitting $\Delta
m_{B_q}$ is defined in terms of $M_{12}^q$
\begin{eqnarray}
\label{defxd}
\Delta m_{B_q}=2\left|M_{12}^q\right|.
\end{eqnarray}
The range of SM predictions for the range of values of $m_t$, $M_W$,
$m_{B_q}$ and $\sqrt {B_{B_q}}f_{B_q}$ are $\Delta
m_{B_d}^{SM}=(2.292$ to $5.318)\times 10^{-10}$ MeV and $\Delta
m_{B_s}^{SM}=(81.991$ to $146.373)\times 10^{-10}$ MeV. Here CKM matrix
element $V_{td}$, $V_{ts}$ and $V_{tb}$ are same as $K_{31}$, $K_{32}$
and $K_{33}$ respectively. We use $V_{td}=0.0079-i0.00347$,
$V_{ts}=-0.03906-i0.000796$ and $V_{tb}=0.9992$ which are obtained
using randomly chosen CKM parameters in the rightmost column in Table
\ref{ckmas}. The range of $\Delta m_{B_s}$ and $\Delta m_{B_d}$ for
the different values of $\beta$ are shown in Table \ref{mdtv} and
\ref{bdat} respectively. We see that the range has a small negative
shift in the negative $\beta$ region and a small positive shift in the
positive $\beta$ region from SM range for $B^0_d-{\bar{B}}^0_d$
system. The feature is reversed for $B^0_s-{\bar{B}}^0_s$ system where
range of $\Delta m_{B_s}$ has a small positive shift in the negative
$\beta$ region and a small negative shift in the positive $\beta$
region from SM range.

Now we want to see the new physics effect on the decay width
difference $\Delta\Gamma_{B_q}$ in $B^0_q-{\bar{B}}^0_q$ system for
both $q=d,s$. The SM results come only from box diagrams \cite{bura1}.
Presence of tree level FCNC coupled to $Z$ and physical Higgs generate
new contributions to the decay width difference in the
$B^0_q-{\bar{B}}^0_q$ system through penguin diagrams. We disregard
Higgs mediated contributions to $M^q_{12}$ which are subdominant
compared to the $Z$ mediated contributions. We now want to calculate
the absorptive part of the amplitude $\Gamma^q_{12}$ for
$B^0_q\rightarrow \bar{B}^0_q$ in this model, keeping only terms linear in
$U_{bq}$. Standard
box, $Z$ and Higgs mediated one loop penguin diagrams give
\begin{eqnarray}
\Gamma^q_{12}=\frac{G_F^2m_b^2
  m_{B_q}(B_{B_q}f_{B_q}^2)}{8\pi}\left[f^q_{SM}(z_c)+U_{bq}\sum_{i=2}^3\xi^q_i\left\{2(f_{\rm box} (i,1)-f_{\rm box} (1,1))+f_{\rm pen}(i)-f_{\rm pen}(1)\right\}\right]\nonumber\\
\label{gamexp}
\end{eqnarray}
where SM result due to box diagrams come through
$f^q_{SM}(z_c)$ \cite{bura1,bura2} whose expression is
\begin{eqnarray}
f^q_{SM}(z_c)=\xi^{q2}_t+\frac{8}{3}\xi^{q}_t\xi^{q}_c(z_c+\frac{1}{4}z_c^2-\frac{1}{2}z_c^3)+\xi^{q2}_c\left\{\sqrt{1-4z_c}(1-\frac{2}{3}z_c)+\frac{8}{3}z_c+\frac{2}{3}z_c^2-\frac{4}{3}z_c^3-1)\right\}.\nonumber\\
\label{smf}
\end{eqnarray}
The absorptive function \cite{bura1} related to SM box is
\begin{eqnarray}
f_{\rm box}(i,j)&=&\frac{1}{3(1-x_i)(1-x_j)}\sqrt{1+(z_i-z_j)^2-2(z_i+z_j)}\times\nonumber\\&&\left[\left\{(1+\frac{x_ix_j}{4})(3-(z_i+z_j)-(z_i-z_j)^2)\right\}+2x_b(z_i+z_j)(z_i+z_j-1)\right]
%\label{}
\end{eqnarray}
whereas the absorptive function related to penguin diagrams mediated by $Z$ and
light Higgs is
\begin{eqnarray}
f_{\rm
  pen}(i)=4\times\frac{-4x_Z\sqrt{1-4z_i}}{3(1-x_i)(x_b-x_Z)}\left[\left\{z_i(g^u_R-g^u_L)(1-\frac{x_b}{4}-\frac{x_i}{2})+\frac{g^u_L}{3}(1+2z_i-\frac{3}{2}x_i)+\frac{g^u_Rx_i}{6}(1+2z_i)\right\}\right . \nonumber\\\left .-\frac{5}{8}\left\{\frac{g^u_L}{3}(1+2z_i)+\frac{x_ig^u_R}{3}(1-z_i)+\frac{x_i}{2x_Z}(1-x_i)+\frac{x_bx_i}{8x_Z}(1+2z_i)(\frac{x_Z-x_H}{x_b-x_H})+\frac{3x_i^2}{4x_Z}(\frac{x_b-x_Z}{x_b-x_H})\right\}\right].\nonumber\\
\label{pen}
\end{eqnarray}
Here $\xi^q_i=V_{ib}^*V_{iq}$, $z_i=m_i^2/m_b^2$, $x_i=m_i^2/M_W^2$,
$x_b=m_b^2/M_W^2$, $x_Z=M_Z^2/M_W^2$, $g^u_L=1/2-2/3\sin^2\theta_W$
and $g^u_R=-2/3\sin^2\theta_W$. $i,j=(1,2,3\equiv u,c,t)$. The
expression $f_{\rm box}(i,j)$ and $f_{\rm pen}(i)$ are only for
$i,j=1,2$. Presence of top inside the loop does not contribute to the
absorptive part due to the kinematical impossibility.  For penguin
diagrams with single up-type quarks inside the loop does not
contribute to the absorptive part. Here the function in Eq.\ 
(\ref{pen}) is for two up-type quarks inside the loop. The $4$ factor
present infront of the expression of $f_{\rm pen}(i)$ in Eq.\ 
(\ref{pen}) is due to the fact that there are four types of diagrams
contributing equally. These are $S$ and $T$ channel diagrams with loop
at the different vertices.  Now the definition of $\Delta\Gamma_{B_q}$
\cite{bura2} is
\begin{eqnarray}
\Delta\Gamma_{B_q}=\frac{4{\rm Re}(M_{12}\Gamma_{12}^{q*})}{\Delta m_{B_q}}.
\label{gama}
\end{eqnarray}
We numerically evaluate the range of
$\left|\Delta\Gamma^{B_q}/\Gamma_{B_q}\right|$ using range of quark
masses, $B^0_q$ masses, decay constants and life times of $B^0_q$.
$m_c=1.15$ to $1.35$ GeV\cite{pdg} where $m_c\equiv {\bar m}_c({\bar m}_c)
$. $\tau_{B_d}=1/\Gamma_{B_d}=(1.536\pm0.014)\times 10^{-12}$s and
$\tau_{B_s}=1/\Gamma_{B_s}=(1.461\pm0.057)\times 10^{-12}$s \cite{pdg}.
Experimental value of the other parameters are given previously. The SM
range of
$\left|\Delta\Gamma^{B_d}/\Gamma_{B_d}\right|=(0.3845$ to $0.9066)\%$ and
$\left|\Delta\Gamma^{B_s}/\Gamma_{B_s}\right|=(12.5191$ to $24.0086)\%$.
The exprimental constraints are
$\left|\Delta\Gamma^{B_d}/\Gamma_{B_d}\right|< 18\%$ at $90\%$ CL \cite{pdg,delphi} and
$\left|\Delta\Gamma^{B_s}/\Gamma_{B_s}\right|<54\%$ at $95\%$ CL
\cite{pdg,gamas}. In our analysis we disregard the masses of $d$, $s$
and $u$ quarks. We also disregard the contributions from heavy scalar
meidiated penguins to $\Gamma^q_{12}$ in Eq.\ (\ref{gamexp}). Their
contributions are suppressed by their mass. We have shown our results
of $\left |\Delta\Gamma^{B_d}/\Gamma_{B_d}\right|$ in Table \ref{bdat}
and $\left |\Delta\Gamma^{B_s}/\Gamma_{B_s}\right|$ in Table
\ref{mdtv} for the different values of the parameter $\beta$. We have
seen that there is small positive shift of decay width in the negative
$\beta$ region and small negative shift in the positive $\beta$ region
for the $B^0_d-\bar{B}^0_d$ system. On the contrary for
$B^0_s-\bar{B}^0_s$ system we have seen that there is small negative
shift of decay width in the negative $\beta$ region and small positive
shift in the positive $\beta$ region.

Single vector like down type quark model has simple expression for CP
asymmetry of $B^0_d$ decay to two channels $J/\psi K_s$ and
$\pi^+\pi^-$. Those are \cite{baren}
\begin{eqnarray}
a_J&\equiv&\frac{\Gamma(B^0\rightarrow
  J/\psi K_s)-\Gamma(\bar{B}^0\rightarrow J/\psi K_s)}{\sin(\Delta m_{B_d}t)(\Gamma(B^0\rightarrow
  J/\psi K_s)+\Gamma(\bar{B}^0\rightarrow J/\psi K_s))}\nonumber\\
&=&\sin(2\beta_0-\arg(\Delta_{bd}))
\label{aj}
\end{eqnarray}
and 
\begin{eqnarray}
a_\pi&\equiv&\frac{\Gamma(B^0\rightarrow
 \pi^+\pi^- )-\Gamma(\bar{B}^0\rightarrow \pi^+\pi^-)}{\sin(\Delta m_{B_d}t)(\Gamma(B^0\rightarrow
 \pi^+\pi^- )+\Gamma(\bar{B}^0\rightarrow \pi^+\pi^-))}\nonumber\\
&=&\sin(2\alpha_0+\arg(\Delta_{bd}))
\label{api}
\end{eqnarray}
where $\alpha_0$ and $\beta_0$ are defined as usual
\begin{eqnarray}
\alpha_0=\arg(-\frac{\xi^d_t}{\xi^d_u})
\label{alpha}
\end{eqnarray}
\begin{eqnarray}
\beta_0=\arg(-\frac{\xi^d_c}{\xi^d_t}).
\label{beta}
\end{eqnarray}
We have calculated the numerical values of those CP asymmetry
parameters $a_J$ and $a_{\pi}$ in this model. The SM value of those
parameters are $a_J=0.7370$ and $a_\pi=0.2549$ using CKM elements with
randomly chosen CKM parameters in Table \ref{ckmas}. The quantity
$\Delta_{bd}$ is defined in Eq.\ (\ref{delbd}). We disregard the
variaton of $\arg(\Delta_{bd})$ with respect to the change of
parameters inside $\Delta_{bd}$ for a particular $\beta$. This
variation is smaller compared to the variation with respect to
$\beta$. So, inspite of getting range we have a value for those CP
asymmetry parameters for a particular $\beta$ value. We show the
results in Table \ref{bdat}. We have seen that
value of the asymmetry parameters have been changed in the third place
and somewhere in the second place after decimal for different values
of $\beta$ from the SM values.
 
Here we cannot have any information about the scale $V$ because of the
cancellations of the scale from the numerator and the denominator in
the second part of $\mathscr{H}$ as Yukawa part dominates over bare
part in the mass of $D$. So the solutions are scale independent. To
know about the scale we look for use of these solutions in strong CP.
\section{Constraints from the Strong CP Phase}
\label{cons}
The existence of topologically nontrivial gauge transformations, and
of field configurations which make transitions between the different
topological sectors of the theory, leads to the existence of the new
term $\Theta \frac{g_s^2}{32\pi^2}F^a_{\mu\nu}\tilde{F}^{a\mu\nu}$ in
$QCD$ action. This term violates parity $P$ and time-reversal $T$. It
will violate $CP$ due to $CPT$ invariance. Diagonalization of the mass
matrix requires different $U(1)$ rotation of the left handed fields
and the right handed fields.  So it effectively generates nonzero
chiral rotation. This modifies the $\Theta$ to $\bar\Theta$, where
$\bar\Theta=\Theta+\arg(\det\bf{m})$, where $\bf{m}$ is quark mass
matrix which appear in ${\bar\Psi}_L {\bf{m}}\Psi_R$, where $\Psi$
contains all quarks. Now the model in which we are working, $CP$ is
invariant in the Lagrangian and breaks only spontaneously. So we cannot keep the $\Theta$ term in our Lagrangian.  The mass terms are
generated through spontaneous breaking of $SU(2)_L\times U(1)_Y$ and
$Z_4$ symmetry. Since $\Theta$ is zero, we can write
$\bar\Theta=\arg(\det\bf{m})$. At the tree level $\det\bf{m}$ is
real and hence $\bar\Theta$ is zero. We should look for one loop
correction of the quark mass matrix. 
Now, loop correction to $\bar\Theta$ is
\begin{eqnarray}
\label{spd}
\delta\bar\Theta &=& \arg\{\det\hskip .1cm ({\bf{m}}-\Sigma)\}\nonumber\\
             &=&-{\bf{Im}}\{{\bf Tr}({{{\bf m}}^{-1}}\Sigma)\},
\end{eqnarray}
where $\Sigma$ is self energy matrix.                     
The one loop correction to the mass matrix has been explicitly done by 
Weinberg in \cite{wein}. Goffin, Segre and Weldon \cite{gof} have
shown that only loops containing physical scalars give nonzero contribution to
the strong  $CP$ phase. Yukawa coupling and mass matrix of the up-type quarks are
real. So they cannot contribute to strong $CP$ phase. Due to this
fact the strong CP phase becomes
\begin{eqnarray}
\label{stcd}
\delta\bar\Theta=-{\bf{Im}}\{{\bf Tr}({{{\mathscr M}}^{-1}}\Sigma)\}.
\end{eqnarray}
where $\mathscr{M}$ is the $4\times 4$ down-type quark mass matrix as in Eq.\
(\ref{mform}) and $\Sigma$ is now the $4\times 4$ down-type quark self energy matrix. In this model
there are three scalar $h$, $s$, and $t$ which are originated from
fluctuation about the vacuum,
$\phi_1$ $\rightarrow$ $v+h$ and
$S$ $\rightarrow$ $\frac{1}{\sqrt{2}}(V+s+it)\exp(i\alpha)$. These
fields are not mass eigenstates. The physical scalars $H_k$ are
related to these scalars $h_a(h,s,t)$ by orthogonal transformations,
$H_k=R_{ka}h_a$. In this basis the Yukawa couplings $\Gamma_a$ of the Lagrangian
 change to $\Gamma_k$, where $\Gamma_k=R_{ka}\Gamma_a$.
The 
contribution of the physical scalars to the self energy
matrix is~\cite{Adhikary:2005je}
\begin{eqnarray}
\label{selfs}
\Sigma^{\phi}&=&-\sum_k\frac{1}{(4\pi)^2}\int_0^1
dx[(1-x)\mathscr{M}\Gamma_k^\dagger+\Gamma_k\mathscr{M}^\dagger]\ln\{\mathscr{M}\mathscr{M}^\dagger 
x^2+M_k^2(1-x)\}\Gamma_k.
\end{eqnarray}  
So the
strong CP phase will be
\begin{eqnarray}
\label{cst1}
\delta\bar\Theta &=&\sum_{k}\frac{1}{(4\pi)^2}\int_0^1 
dx{\bf{Im}}[{{\bf{Tr}}}\{\mathscr{M}^{-1}\Gamma_k
\mathscr{M}^\dagger
\ln(\mathscr{M}\mathscr{M}^\dagger 
x^2+M_k^2(1-x))
\Gamma_k\} ].
\end{eqnarray}
We can write the strong CP phase in terms of original Yukawa couplings
of the Lagrangian. Changing also the logarithmic part of Eq.\
(\ref{cst1}) in the diagonal basis the dominant part of the strong CP
phase comes from $D$ quark mass. Then the strong CP phase will be 
\begin{eqnarray}
\label{cst2}
\delta\bar\Theta &=&\sum_{k,a,b}\frac{R_{ka}R_{kb}}{(4\pi)^2}\int_0^1 
dx{\bf{Im}}[\mathscr{U}^\dagger\Gamma_a \mathscr{M}^{-1}\Gamma_b
\mathscr{M}^\dagger\mathscr{U}]_{44}
\ln({M_D}^2 
x^2+M_k^2(1-x)).
\end{eqnarray}
The above expression gives a nonzero value only when $a=t$ and $b=h$.
The explicit calculations of strong CP phase have been done earlier
\cite{Branco1}. To
keep the hierarchy between the scales $v$ and the scale $V$ and to
avoid the fine tuning in the stationarity equations of the scalar
potential in Eq.\ (\ref{scalp}) we should consider two
parameters $c_1$ and $c_2$ of the scalar potential small
as $O(v^2/V^2)$. Light scalar mass $M_1$ remains of the order of $v$.
The other two scalar masses become 
$M_2$$\sim$$M_3$$\sim$$V$. So the calculation of the order of
magnitude of $R_{kt}R_{kh}$$\sim$$2\sin(2\alpha)\frac{v^3}{V^3}$\cite{Branco1}. In performing integrations we keep the
assumption $M_{2,3}$ $\ge$$2M_D$ same as in \cite{Branco1} but
$2M_D\ge M_1$ is not valid for all $\beta$ as in Table \ref{mdtv} and
as we consider $M_1=M_H=150$ GeV. Under the considerations $M_2$
and $M_3$ are nearly same and 
$M_{2,3}$$\gg$$2M_D$ we get the following expression of strong CP phase
 \begin{eqnarray}
\label{strp}
\delta\bar\Theta \approx\frac{\sin{2\alpha}}{16\pi^2}
({\bf f}_q^2-{\bf f^\prime}_q^2)\frac{v^2}{V^2}
\left[1+\ln\left(\frac{V^2}{M_D^2}\right)-I_1\right]  
\end{eqnarray} 
where\\
\begin{eqnarray}
I_1 = \cases{
\frac{M_1^2}{M_D^2}\left[\frac{1}{2}\ln
  \left(\frac{M_1^2}{M_D^2}\right)+\left(\frac{4M_D^2}{M_1^2}-1\right)^{\frac{1}{2}}
\tan^{-1}\left(\frac{4M_D^2}{M_1^2}-1\right)^{\frac{1}{2}}\right]
& \mbox{for $2M_D\ge M_1$,} \cr
\frac{M_1^2}{M_D^2}\left[\frac{1}{2}\ln
  \left(\frac{M_1^2}{M_D^2}\right)-
\left(1-\frac{4M_D^2}{M_1^2}\right)^{\frac{1}{2}}
\tanh^{-1}\left(1-\frac{4M_D^2}{M_1^2}\right)^{\frac{1}{2}}\right]
& \mbox{for $2M_D\le M_1.$}
}
\label{i1}
\end{eqnarray}
%%
%%
%%%%
%%%%% 
\begin{table}[t]
\begin{center}
\begin{tabular}{|c|r|c|c|c|r|}
\hline
$\beta$ &   Lower Bound & Upper bound & Range of & Range of &  Lower Bound \\
 &    of  $M_D$[GeV]  &   of $\left |U_{bs}\right |$ $\times$ $10^{5}$
 & $\Delta m_{B_s}\times 10^{10}$[MeV] &
 $\left |\frac{\Delta\Gamma_{B_s}}{\Gamma_{B_s}}\right |\times 10^2$ & of $V$ [TeV] \\
\hline
$-10.0$  & $59.1$ & $0.838$&  $82.038$ to $146.459$ & $12.5139$ to $23.9984$ &$10.1$\\
\hline
$-7.5$   & $72.9$&  $0.695$& $82.033$ to $146.449$ & $12.5145$ to $23.9995$ &$12.9$\\
\hline
 $-5.0$  & $108.1$&  $0.506$& $82.023$ to $146.432$& $12.5155$ to $24.0016$ & $17.4$\\
\hline
$-4.5$  & $123.4$&  $0.461$& $82.021$ to $146.427$& $12.5158$ to $24.0022$ &$19.1$\\
\hline
$-4.0$   & $146.6$&  $0.414$& $82.018$ to $146.422$& $12.5161$ to $24.0028$ & $21.6$\\
\hline
$-3.5$  & $187.5$&  $0.364$& $82.015$ to $146.417$& $12.5165$ to $24.0034$ & $25.3$\\
\hline
$-3.2$  & $234.2$&  $0.333$& $82.013$ to $146.413$& $12.5167$ to $24.0038$ & $29.0$\\
\hline
$2.2$   & $272.8$&  $0.379$& $81.965$ to $146.328$& $12.5219$ to $24.0140$ & $30.5$\\
\hline
$2.5$   & $191.3$&  $0.421$& $81.963$ to $146.324$& $12.5222$ to $24.0146$ & $24.7$\\
\hline
$3.0$   & $136.2$&  $0.493$& $81.958$ to $146.316$& $12.5226$ to $24.0155$ & $19.9$\\
\hline
$3.5$  & $108.8$&  $0.563$& $81.955$ to $146.309$& $12.5231$ to $24.0163$ & $17.1$\\
\hline
 $4.0$  & $91.9$&  $0.631$& $81.951$ to $146.302$& $12.5235$ to $24.0171$ & $15.1$\\
\hline
$4.5$  & $80.2$&  $0.697$& $81.947$ to $146.297$& $12.5238$ to $24.0178$ & $13.7$\\
\hline
$5.0$  & $71.6$&  $0.761$& $81.945$ to $146.292$& $12.5241$ to $24.0184$ & $12.6$\\
\hline
$7.5$   & $49.7$&  $1.034$& $81.937$ to $146.278$& $12.5249$ to $24.0200$ & $9.3$\\
\hline
$10.0$   & $41.2$&  $1.220$& $81.938$ to $146.280$& $12.5248$ to $24.0197$ &$6.9$\\
\hline
\end{tabular}\\
\caption{\label{mdtv} The lower bound on $M_D$
  using the obtained upper bound of $U_{bb}\le 7.204\times 10^{-4}$
  for the different values of $\beta$. The upper bound of  $U_{bs}$,
  range of  $\Delta m_{B_s}$, range of  $\left
    |\frac{\Delta\Gamma_{B_s}}{\Gamma_{B_s}}\right |\times 10^2$ and the lower bound of the scale $V$ for
  the different values of $\beta$. We use the solutions for the inputs
  which are in the rightmost column of Table \ref{ckmas}. We use
  $\alpha(M_Z)=1/128.91$, 
$\alpha_s(M_Z)=(0.1187\pm0.002)$, $M_Z=(91.187\pm0.0021)$ GeV,
$M_W=(80.425\pm0.038)$ GeV, $M_H=150$ GeV, $\sin^2
\theta_W=1-M_W^2/M_Z^2$, $m_t=(174.3\pm5.1)$ GeV and $m_b=4.1$ to $ 4.4$ GeV
\cite{pdg}. }  
\end{center}
\end{table}
%%
%%
%
%%%%
%%%%% 
\begin{table}[t]
\begin{center}
\begin{tabular}{|c|c|c|c|c|c|}
\hline
$\beta$ &   Upper bound & Range of  & Range of & $a_J$ for & $a_\pi$ for\\
 &       of $\left |U_{bd}\right |$ $\times$ $10^{5}$
 & $\Delta m_{B_d}\times 10^{10}$[MeV]
 &$\left |\frac{\Delta\Gamma_{B_d}}{\Gamma_{B_d}}\right |\times 10^2$ & ${B_d\rightarrow J/\psi
   K_s}$  & ${B_d\rightarrow\pi^+\pi^-}$  \\
\hline
$-10.0$  &   $3.652$& $2.273$ to $5.274$& $0.3864$ to $0.9108$ & $0.7434$&$0.2641$\\
\hline
$-7.5$   &  $3.027$& $2.275$ to $5.279$& $0.3863$ to $0.9108$ & $0.7418$&$0.2615$\\
\hline
 $-5.0$  &   $2.204$& $2.278$ to $5.287$&  $0.3861$ to $0.9102$ & $0.7403$&$0.2597$\\
\hline
$-4.5$  &  $2.009$& $2.279$ to $5.289$&  $0.3859$ to $0.9100$ & $0.7400$&$0.2592$\\
\hline
$-4.0$   &  $1.802$& $2.280$ to $5.291$& $0.3859$ to $0.9097$ & $0.7396$&$0.2587$\\
\hline
$-3.5$  &   $1.585$& $2.282$ to $5.294$&  $0.3858$ to $0.9095$ & $0.7392$ & $0.2581$\\
\hline
$-3.2$  &   $1.449$& $2.282$ to $5.295$& $0.3857$ to $0.9093$ & $0.7390$&$0.2578$\\
\hline
$2.2$   &  $1.649$& $2.307$ to $5.353$& $0.3829$ to $0.9026$ & $0.7359$&$0.2534$\\
\hline
$2.5$   &   $1.836$& $2.309$ to $5.358$& $0.3827$ to $0.9022$ & $0.7360$&$0.2535$\\
\hline
$3.0$   &  $2.146$& $2.312$ to $5.365$& $0.3824$ to $0.9013$ & $0.7361$&$0.2536$\\
\hline
$3.5$  &   $2.451$& $2.316$ to $5.373$& $0.3820$ to $0.9005$ & $0.7363$&$0.2540$\\
\hline
 $4.0$  &   $2.748$& $2.319$ to $5.381$&  $0.3817$ to $0.8997$ & $0.7367$&$0.2545$\\
\hline
$4.5$  &   $3.037$& $2.323$ to $5.388$& $0.3814$ to $0.8990$ &  $0.7371$&$0.2551$\\
\hline
$5.0$  &  $3.316$& $2.326$ to $5.396$& $0.3811$ to $0.8982$ & $0.7377$&$0.2559$\\
\hline
$7.5$   &   $4.507$& $2.338$ to $5.423$&  $0.3798$ to $0.8952$ & $0.7416$&$0.2615$\\
\hline
$10.0$   &  $5.317$& $2.341$ to $5.431$& $0.3792$ to $0.8936$ & $0.7458$&$0.2676$\\
\hline
\end{tabular}\\
\caption{\label{bdat} The upper bound of  $U_{bd}$,
  range of  $\Delta m_{B_d}$, range of $\left
    |\frac{\Delta\Gamma_{B_d}}{\Gamma_{B_d}}\right |\times 10^2$, CP asymmetry parameters $a_J$ and
  $a_\pi$ for the decay channels $B^0_d\rightarrow J/\psi K_s$ and
  $B^0_d\rightarrow \pi^+\pi^-$ respectively for
  the different values of $\beta$. We use the solutions for the inputs
  which are in the rightmost column of Table \ref{ckmas}. $M_D$ values
  are taken from the Table \ref{mdtv} to calculate $U_{bd}$ bound. We
  use also the inputs in the caption of Table \ref{mdtv}.
\cite{pdg}. }  
\end{center}
\end{table}
The integrals are performed by Bento, Branco and Parada in \cite{Branco1}. Using our ansatz of the mass matrix
 of down-type quarks we have strong
CP phase of the following form
\begin{eqnarray}
\label{sedl}
\delta\bar\Theta\approx\frac{\sin{2\alpha}}{8\pi^2}\Delta{f}({\Delta{f}+2{f}})\frac{M_D^2v^2}{V^4}\left[1+\ln\left(\frac{V^2}{M_D^2}\right)-I_1\right].  
\end{eqnarray}
The strong phase  will vanish for
$\Delta$${f}$=$0$, $\alpha$=$0$, $\pi/2$, $\pi$, $2\pi$. Weak phase will
also vanish for those values of $\Delta f$ and $\alpha$ because
for each value ${M_D^0}^\dagger M_D^0$ and hence $\mathscr{H}$ become
real symmetric. In terms of $x$
parameters the form of strong CP phase using the relations in Eq.\
(\ref{fxr}) we have
\begin{eqnarray}
\label{stx}
\delta\bar\Theta\approx\frac{1}{4\pi^2}x_7x_8x_9\frac{M_D^2v^2}{V^4}\left[1+\ln\left(\frac{V^2}{M_D^2}\right)-I_1\right].  
\end{eqnarray}
 Present value of the experimental bound on electric dipole moment of neutron
 gives value of upper bound  on strong CP phase, $\delta\bar\Theta\le 2\times
 10^{-10}$. We can convert the bound of $\delta\bar\Theta$ to bound of
 the scale $V$ using the Eq.\ (\ref{stx}), the solutions of the Eq.\
 (\ref{emd}) and
 the bound of $M_D$ in Table \ref{mdtv}. We show the lower bounds on the scale
 $V$ for the different values of $\beta$ in Table \ref{mdtv}. The bounds on the
 scale $V$ for the different sets of solution for the given inputs
 will differ within $1$ TeV.
 There exists a tiny window on the scale
$V$ for the bound of $\delta\bar\Theta$ for the expression in Eq.\
(\ref{stx}) which is small like $v$. We discard
this result because the result come for fine cancellation among
the terms inside the parenthesis in Eq.\
(\ref{stx}). 

\section{Conclusion}
\label{conclu} 
The model of BPR \cite{Branco2} claims that all CP violating phenomena
can originate from a single phase which appears in the VEV of SM gauge
singlet complex scalar $S$. Bento, Branco and Parada \cite{Branco1}
started this work where they showed that the weak CP phase and the
strong CP phase can have a common origin. BPR \cite{Branco2}
extend this to the leptonic sector. 

We have made quantitative studies of this model in the quark sector.
We find the Lagrangian parameters for a specific kind of ansatz of
down-type quark mass matrix where CKM phase is generated minimally.
The Lagrangian of the BPR model is CP invariant.  CP is broken
spontaneously through a single phase $\alpha$ in the VEV of the
singlet scalar $S$. Hence, all Lagrangian parameters are real and the
phases are only due to $\alpha$. We observe that in the interval
$-3.2<\beta<2.12$ the ansatz of Eq.\ (\ref{fdf}) becomes inconsistent.
We find the numerical value of the Lagrangian parameters outside this
interval. We also find that the contribution of the bare part to the
mass of $D$ quark is negligible compare to the Yukawa part. This
observation is independent of $\beta$. One point we should make about
the unitarity of $\mathscr{U}$ Eq.\ (\ref{uform}).  To calculate $U$
we use the solution of the Eq.\ (\ref{emd}) where $K$ is unitary.  We
calculate approximate $R$, $S$ and $T$ of Eq.\ (\ref{ulbs}) and Eq.\ 
(\ref{sfm}) with the solutions Set-1 for $\beta=3.5$ in Table
\ref{xsol} and the value of $M_D$ for the same solution in Table
\ref{mdtv}.  Putting these $R$, $S$ and $T$ and unitary $K$ in
$\mathscr{U}$ Eq.\ (\ref{uform}) to test its unitarity we see that
${\mathscr U}{\mathscr U}^\dagger $ and ${\mathscr U}^\dagger{\mathscr
  U}$ deviate from identity by the amount only $O(10^{-4})$ for a few
elements. This deviation is not so sensitive to the parameter $\beta$.

We have observed that this kind of vector quark model cannot increase
the partial decay width of $Z\rightarrow \bar b b$. So we find out the
bound on $U_{bb}$ from the condition that the new physics decrease the
SM value up to experimental lower bound of $\Gamma_{Z\rightarrow \bar
  b b}$ considering $1\sigma$ error of the total decay width of $Z$
and the branching ratio of this decay mode of $Z$.  Here we also
should point out about value of $M_H$. Suppose we fix the value of
$\beta$ at $3.5$ and change $M_H$ from $150$ GeV to $100$ GeV. Then
the decay width of $Z\rightarrow \bar b b$ decreases by nearly $0.01$
MeV, $M_D$ remains almost same whereas $V$ increases by nearly $0.34$
TeV. On the other hand, for the same value of $\beta$, if $M_H$ is
changed from $150$ GeV to $300$ GeV, it increases the decay width of
$Z\rightarrow \bar b b$ by nearly $0.007$ MeV, $M_D$ remains almost
same whereas $V$ decreases by nearly $0.62$ TeV. Scope of new physics
remains in determining the mass difference $\Delta m_{B_q}$ for both
$q=d,s$.  Here we have only small shift of the $\Delta m_{B_q}$ range
from SM range. Actually huge uncertainty of $\sqrt {
  B_{B_d}}f_{B_d}=221\pm 28^{+0}_{-22}$ and $\sqrt {
  B_{B_s}}f_{B_s}=255\pm 31$ introduce huge uncertainty in theoretical
value of $\Delta m_{B_d}$ and $\Delta m_{B_s}$ respectively.  This
model has very small effect on decay width difference in both
$B^0_d-\bar{B}^0_d$ and $B^0_s-\bar{B}^0_s$ system.  The results are
shown in Table \ref{bdat} and \ref{mdtv} respectively.  We also have
observed in Table \ref{bdat} that the numerical values of the CP
asymmetry parameters $a_J$ and $a_\pi$ for the decays
$B^0_d\rightarrow J/\psi K_s$ and $B^0_d\rightarrow \pi^+ \pi^-$
respectively have effect on the third and somewhere on the second
place after the decimal from SM value. This part of our analysis is quite
general and applies to any model containing extra down-type quarks,
e.g., models inspired by $E_6$ grand unification.

The strong CP phase in this model is suppressed by inverse powers of
$V$. It should be noted that we have used the solutions of the
elements of the quark mass matrix to obtain lower bounds on $V$.
There are, however, direct experimental limits on the mass of strongly
produced massive quarks of charge $-1/3$.  This lower bound of $199$
GeV \cite{aff} is derived for a fourth generation down-type quark
which is produced strongly in pair and decay to $bZ$ via 1-loop FCNC.
If this bound is assumed to hold for $M_D$ in the BPR model where FCNC
exists at the tree level, the lower bound on $V$ increases.  For
example, with $\beta=-10$, we now get $V>20.3$ TeV, whereas for
$\beta=+10$, we get $V>15.4$ TeV.

Note added: After completion of our work a new analysis of a group was
published \cite{buch}, where the experimental value of $m_b$ has
changed to a larger value with a smaller error bar,
$m_b=4.591\pm0.040$ GeV in the kinetic scheme at scale $1$ GeV. But we
have used the PDG \cite{pdg} value $m_b=4.1$ to $4.4$ in our numerical
analysis in this model, and the random value chosen in Table
\ref{ckmas} for finding Lagrangian parameters lies outside the
$1\sigma$ range advocated in \cite{buch}.

\paragraph*{Acknowledgments}
The author acknowledges helpful
discussions with Palash B. Pal and Gautam Bhattacharyya at various
stages of the work. He also thanks them for reading the manuscript and
suggesting improvements.

%%%%%%%%%%%%%%%%%%%%

\end{document}